# Integrated Transmission & Distribution System Modeling and Analysis: Need & Advantages

Himanshu Jain, Kaveh Rahimi, Ahmad Tbaileh, Robert P. Broadwater

Electrical and Computer Engineering
Virginia Tech
Blacksburg, VA, USA - 24061
jhim86@vt.edu, krahimi@ieee.org,
atahm12@vt.edu, dew@vt.edu

Akshay Kumar Jain

Electrical Engineering
Indian Institute of Technology, Roorkee
Roorkee, Uttarakhand, India - 247667
akshaykj92@gmail.com

Murat Dilek

Electrical Distribution Design (EDD)
Blacksburg, VA, USA – 24060
murat-dilek@edd-us.com

*Abstract*—The primary objective of this paper is to highlight the need for and benefits of studying the steady state and dynamic response of power systems using three phase integrated transmission and distribution (T&D) system models (hereafter referred to as hybrid models). Modeling and simulation requirements for building and analyzing hybrid models are also discussed. Finally, results from steady state and dynamic simulations of a large hybrid model are presented to demonstrate insights that can be obtained from hybrid models which cannot be obtained from the study of transmission and distribution systems separately.

*Index Terms*-- distributed power generation, power distribution, power system dynamics, power system simulation, smart grids

## I. Introduction: Need for Hybrid Models

Power systems are typically divided into high voltage transmission (typically 69 kV and above) and low voltage distribution systems for planning and operations. The sheer size of power systems, jurisdiction of state and federal bodies on various aspects of their planning and operations, and the hitherto reliable operation of separately designed T&D systems are some of the factors due to which the planning and operations of power systems as a single entity has not received much attention. However, it is time that the *separated approach* for planning and operating power systems be revisited because of reasons that will now be considered.

**Asymmetry of Transmission Lines**: The conventional approach for studying steady-state and dynamic response of the bulk power system is to assume that the transmission network is balanced. One of the driving factors behind this assumption is the presumed symmetry of transmission lines due to transposition. However, many high voltage transmission lines are not transposed [1], [2] and assuming transposition for such lines during planning and operations may cause adverse events to be missed. [2] lists several instances when asymmetry of transmission lines resulted in adverse impacts on the power system such as sequential tripping of generators due to excessive negative sequence voltage and extensive damage to customer rotating machinery.

**Decentralized Power Generation**: Distributed generation (DG), particularly solar photovoltaic (PV) based DG, connected to the distribution system is gradually reaching a penetration level at which its impact on the transmission system cannot be overlooked. Fig. 1 obtained from [3] shows the expected rapid growth in solar PV based DG over the next decade. As the penetration of DG increases, its impact on the steady state and dynamic response of power systems needs to be thoroughly reviewed. The following excerpt from [4] echoes this concern:

*"Interviews with IOU [Investor Owned Utilities] engineers identified the need to conduct studies to determine the level of penetration [of variable generation] that may adversely impact such issues as system stability"*

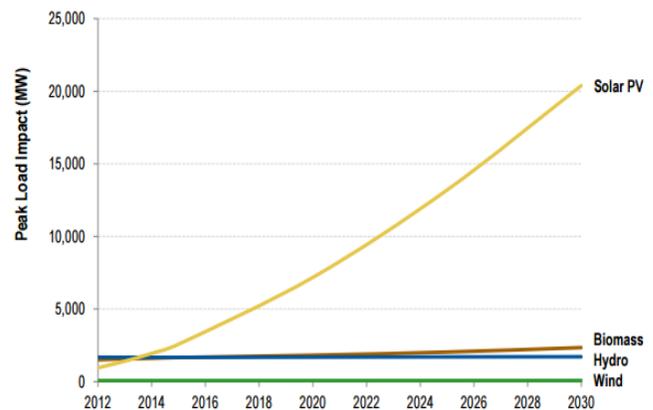

Figure 1. Projected impact of distributed generation at peak load in eastern interconnection [3]

Besides displacing central power plants, single phase rooftop solar PV based DG can aggravate imbalance in the power system. The imbalance can result from asymmetrical placement of DG on the three phases, and cloud movement that may not affect DG placed on different phases identically. The potential for increased imbalance due to DG is highlighted in the following excerpt from [4]:

*"These [procedures used by transmission operators] would need to be expanded to include how to manage imbalances caused by increased DG on the system as this would be a new phenomenon"*

To model the asymmetry of transmission lines, three phase modeling and simulation of transmission networks is required. Similarly, accurate assessment of the impact of DG on power systems would require combined models of transmission and distribution networks. Both these requirements can be met by developing hybrid models of power systems and software that can simulate their steady state and dynamic response. Therefore, section II discusses the key features of hybrid models and the software needed to analyze them. In section III, results of simulating the steady state and dynamic response of a hybrid model using algorithms developed by the authors are discussed to demonstrate the insights about power systems behavior that are revealed by hybrid models.

## II. BUILDING AND ANALYZING HYBRID MODELS

Two key ingredients are required for building hybrid models: (i) three phase equipment models (e.g., transformers, transmission lines, distributed generation); and (ii) three phase network topology of T&D networks. Once the hybrid model is built, software that can simulate its steady state and dynamic response is needed. This section discusses the challenges, and potential ways of meeting the challenges, in building and analyzing hybrid models.

### A. Three Phase Equipment Models

Steady state and dynamic models for conventional power system equipment, such as transmission lines, transformers, voltage regulators, synchronous and induction machines, can be found in several textbooks and research papers [5]-[7]. The models for these equipment are largely standardized and will not be discussed here. Instead, focus will be on modeling of inverter based DG as solar PV based DG is expected to dominate the DG portfolio in the foreseeable future.

To study the steady state behavior of hybrid models, solar PV based DG can be interfaced with the network as a controlled current source [8]. Modeling inverter based DG for studying the dynamic response of hybrid models is, however, more difficult. Inverter based DGs are different from generators that interface with the network through electric machines. While the interaction of electric machines with the network is governed by their physics, interaction of inverter based DGs is governed by the controller built into the inverter. Since the controllers are proprietary, it is difficult to obtain generic dynamic models for inverter based DGs which makes coordination with inverter vendors and utilities important for building accurate dynamic models.

In recent years, the Western Electricity Coordinating Council (WECC) has developed generic dynamic models for large solar PV power plants [9]. However, these models are recommended for use under balanced network conditions and it is assumed that the solar power input to the plant does not change over the study duration which is about 20-30 seconds. The National Renewable Energy Laboratory (NREL) has also published several reports and research papers that can be used as guidelines for developing dynamic models of three phase and single phase inverter based DGs [8], [10], [11]. An interesting observation from both [11] and [12] is that when the objective of a study is to assess the impact of solar PV based DG on power system dynamic response, an average model for solar PV inverter may suffice. This property greatly simplifies the dynamic models of inverter based DG as the high frequency switching dynamics of inverters do not need to be modeled.

### B. Three Phase Network Topology of T&D Networks

Many programs (e.g., DEW, CYME, EMTP-RV, Gridlab-D, Open DSS and ATP) have the capability to model three phase network topologies. Predominantly, three phase network topologies are built for distribution networks since transmission networks are assumed to be balanced. As discussed earlier, asymmetry of transmission networks must also be included in hybrid models.

References [13]-[15] discuss an approach for developing hybrid models where such models are automatically built from multiple data sources available within a utility. Manual intervention for developing the models is kept to a minimum. To the best of our knowledge, the hybrid models developed using this approach are assumed to be balanced. Later in the paper an Integrated System Modeling (ISM) approach for building and analyzing hybrid models is discussed. Hybrid models built using the ISM approach can be unbalanced and include equipment with one, two or three phases.

### C. Software for Simulating Steady State and Dynamic Response of Hybrid Models

Simulating the steady state response of power systems (power flow analysis) using hybrid models is challenging because numerical methods that are stable for transmission networks, do not work well for distribution networks [7], [16]. Therefore, power flow analysis methods that can solve both transmission and distribution networks together are needed. [17] discusses one such method which is used in the Distributed Engineering Workstation (DEW) software.

Calculating the dynamic response of hybrid models is also challenging. Dynamic response simulation of balanced transmission networks is performed by solving a set of Differential Algebraic Equations (DAEs), where the components with fast dynamics, such as transmission lines, are modeled using algebraic equations (frequency domain), while components with slower dynamics, such as synchronous machines and their controllers, are modeled using differential equations (time domain) [5], [6], [18]. Solving a similar set of DAEs under network imbalance becomes challenging because at the interface of time and frequency domains, six phasor variables (three magnitudes and three angles) should be calculated from three time domain values. References [19]-[21] discuss this problem in detail and present a method for solving the problem which has been implemented in the Three Phase Dynamics Analyzer (TPDA) application of DEW.

Electromagnetic Transient Simulation Programs (EMTP), such as EMTP-RV and the Alternative Transients Program (ATP), can also be used to simulate the dynamic response of hybrid models. These programs model the entire power system using differential equations. Reference [22] discusses the application of an EMTP program for simulating the transient response of an actual distribution circuit in New York. However, we do not know of any application of EMTP programs for simulating the dynamic response of hybrid models, which may contain hundreds of distribution circuits besides the transmission network.

### D. Integrated System Modeling (ISM) Approach

In order to build an accurate hybrid model of a power system it is critical that data and models are seamlessly shared between transmission and distribution domains. Moreover, various algorithms that use the hybrid model should also be able to easily share their results. For example, a Monte Carlo method based reliability analysis application should be able to call the power flow analysis application repeatedly for accurately calculating the load unserved due to probabilistically generated contingencies. The ISM approach introduced in [23] enables seamless sharing of data and analysis results.

In the ISM approach, a common object oriented modeling architecture is made available to various utility departments to build and maintain equipment and network topology models. Changes made to the models by a department immediately become available to all the analysis algorithms. For example, in a utility that uses the ISM approach to model their power system, changes made to the configuration of transmission or distribution lines can be added to the model by the concerned department and the impact of modified impedances on power flow can be immediately evaluated by the power flow analysis application being used by the operations department. Using the ISM approach a utility can build and maintain accurate and up-to-date hybrid models of their network, and use these models for more efficient planning and operations.

## III. SIMULATING THE STEADY STATE AND DYNAMIC RESPONSE OF A HYBRID MODEL

The primary focus of this section is to demonstrate the advantages of using hybrid models to study the steady state and dynamic response of power systems. The simulations are performed using power flow and dynamic simulation applications (TPDA) of DEW. Detailed discussion on the power flow analysis application used in DEW can be found in [17] while the theory behind TPDA, its key features, and validation with PSLF and ATP can be found in [19]-[21].

### A. Description of the Hybrid Model

The hybrid model used in the paper consists of the IEEE 39 bus transmission system [24] whose 12 of the 19 loads are modeled using the IEEE 123 bus distribution feeders [25]. At each of these 12 transmission load buses, the lumped load is replaced by a distribution circuit that contains 8 IEEE 123 bus distribution feeders. Fig. 2 shows the substation configuration used in the hybrid model while Table 1 lists the number and types of components in the hybrid model. More details about the hybrid model can be found in [26].

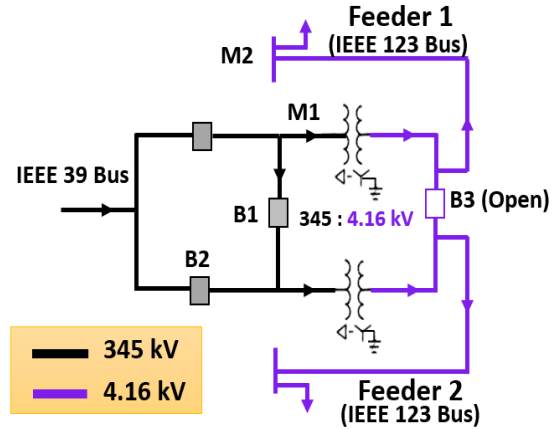

Figure 2. Substation configuration used in the hybrid model

TABLE I
COMPONENTS OF THE HYBRID MODEL

| Component | Quantity | Component | Quantity |
|---|---|---|---|
| Distribution Feeders | 96 | Breakers/Switches | 1,154 |
| 1-Phase Lines | 5,280 | 1-Phase Loads | 7,584 |
| 2-Phase Lines | 288 | 2-Phase Loads | 96 |
| 3-Phase Lines | 5,280 | 3-Phase Loads | 500 |
| 3-Phase Underground Cables | 480 | Voltage Regulators | 384 |
| Transformers | 134 | Shunt Capacitors | 384 |

### B. Steady State Simulation (Power Flow Analysis)

DEW uses two approaches for building hybrid models, and detailed substation models that tie transmission, sub-transmission, and distribution networks together can be included in models built using either approach as shown in Fig. 3. Also shown in Fig. 3 is the difference in interaction during the power flow between the T&D networks of the hybrid models built using the two approaches. A brief description of the two approaches is given below.

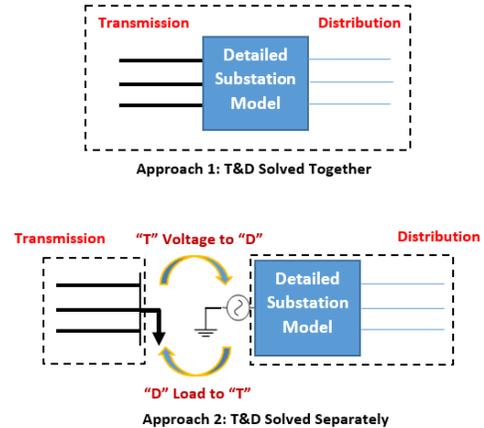

Figure 3. Illustration of two approaches used in DEW to build and simulate hybrid models

In one modeling approach transmission lines come directly into the substation model and distribution lines leave. This modeling approach is very direct and corresponds to how the system is actually built. Hybrid models built using this approach are solved as one large system by the power flow application of DEW.

In the second modeling approach there is no physical connection between the transmission and distribution systems, and they interact through three phase transmission system load buses that are tied to three phase distribution system sources as shown in Fig. 3. The steady state solution of the hybrid models built using this approach begins with the power flow solving all the distribution systems by assuming nominal voltages at their distribution sources. Once the distribution systems are solved, the power flowing through the distribution sources are assigned to the corresponding transmission load buses. Next, the transmission system is solved and voltages at the transmission load buses are assigned to the corresponding distribution sources for solving the distribution systems in the next iteration. Therefore, for the models built using the second approach, the transmission and distribution systems are iteratively solved separately, with results being exchanged by boundary components (i.e., transmission load buses and distribution sources) till convergence criteria are met.

In DEW the second modeling approach has the advantage of easier extraction of distribution circuits from the hybrid model, as each circuit has its own source. It has also been observed that, in general, the power flow converges faster in hybrid models built using the second approach. For example, while the power flow converged in 16 seconds in the hybrid model of a utility that was built in DEW using the first approach, convergence was achieved in 4 seconds when the model was built using the second approach. However, for the hybrid model described in section III.A, power flow converged in about 5 seconds for both modeling approaches. The results of power flow analysis of this hybrid model are discussed below.

Fig. 4 plots the voltages of three phase buses (voltage imbalance increases from left to right) in one feeder of the distribution circuit which is supplied by transmission bus 18 of the hybrid model (same as bus 18 of the IEEE 39 bus transmission system).

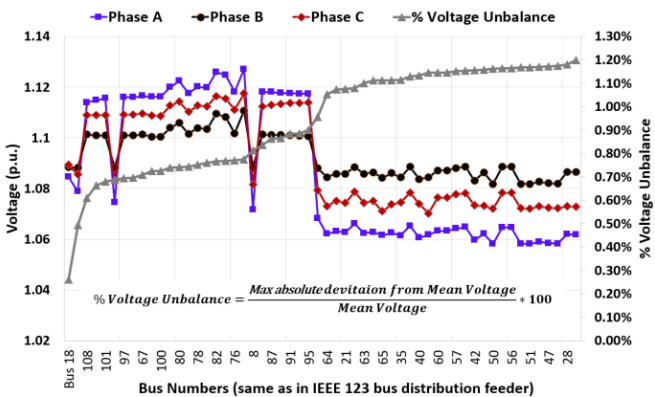
Figure 4. Three phase bus voltages in a feeder of the distribution circuit supplied by transmission bus 18

The following observations can be made from Fig. 4 which highlight the advantages of using hybrid models for steady state analysis of power systems:

a.) The transmission bus voltage is almost balanced with only 0.26% imbalance. On the other hand, minimum voltage imbalance in the distribution buses is 0.49%, and almost 50% of them have greater than 1% voltage imbalance. A transmission only model would not have revealed this variation in voltage imbalance in the distribution feeder.

b.) Hybrid models improve the efficiency of steady state analysis. Insights about voltage imbalance at the transmission and distribution buses were obtained in a single step without the need for separate analysis of transmission and distribution systems using different simulation platforms.

C. *Dynamic Simulation*

To simulate the dynamic response of the hybrid model, the approach discussed in [19] and [21] is used. The 10 synchronous generators connected to the transmission system are modeled using the GENROU model, while their exciters are modeled using the ST1A model. The parameters used for these dynamic models can be found in [26].

For simulating the transient response of the hybrid model an 18 ohm single line to ground fault on Phase A was initiated at bus 12 (same as bus 12 of the IEEE 39 bus transmission system) of the hybrid model, slightly right to the breaker B2 of Fig. 2 (bus 12 has the same substation configuration as shown in Fig. 2). The fault was initiated at the end of the 10$^{th}$ cycle and was cleared after 10 cycles by opening breakers B1 and B2. Voltages at locations M1 (bus 12) and M2 (2.4 kV single phase load in one of the feeders supplied by bus 12) of Fig. 2 were recorded.

Fig. 5 plots the phase A voltages at M1 and M2 along with the under voltage region of the ITI curve which is drawn based on [27].

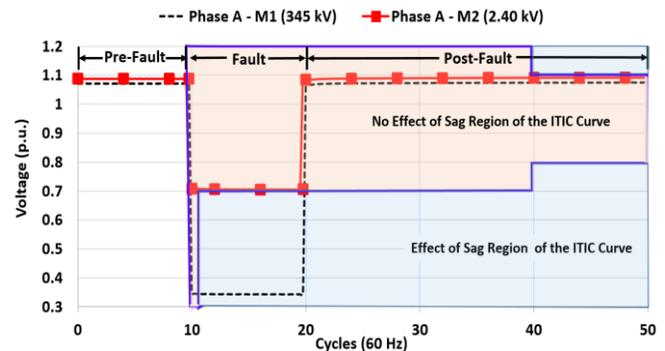
Figure 5. Phase A voltage sag at bus 12 & at a distribution load supplied by bus 12

From Fig. 5 it can be seen that while the transmission bus voltage suggests that loads connected to Phase A of bus 12 can experience power quality issues such as stalling of air conditioners, and restarting of computers as the voltage remains below 0.4 p.u. for 10 cycles [27], the distribution bus voltage suggests that no such problem should be encountered by the loads, as the voltage is within the "no effect of sag" region of the ITI curve. While not shown here, all the distribution bus voltages were found to be within the "no effect of sag" region of the ITI curve.

These observations show the benefits that utility engineers can derive by performing dynamic simulations using hybrid models. In the present case utility engineers will not need to spend time and resources for addressing the power quality problems that the transmission only model would have flagged, but which were found to be non-existent by the hybrid model.

IV. CONCLUSION

The primary objective of this paper was to highlight the need for and benefits of studying the behavior of power systems using hybrid models. Challenges that may be encountered in building and analyzing hybrid models were discussed, and potential ways for meeting these challenges were identified. In the last section the steady state and dynamic response of a large hybrid model that contained over 10,000 lines and over 8,000 loads was simulated. The simulation results demonstrated that interesting insights into the behavior of power systems can be easily obtained by analyzing hybrid models which would be difficult, if not impossible, to obtain by studying transmission and distribution systems separately.

The authors are currently working on studying the impact of high penetration distributed generation, particularly solar PV based DG, on the power system of a utility using the hybrid model of their system. The authors plan to share the results of this study with the power systems community in the near future.